\documentstyle[aps]{revtex}
\begin{document}

\title{Quantum orientational melting of mesoscopic clusters}

\author{
A. I. Belousov and Yu. E. Lozovik
}

\address{Institute of Spectroscopy, Russian Academy of Sciences,
142092 Troitsk, Moscow region, Russia}

\maketitle

\begin{abstract}
By path integral Monte Carlo simulations
we study the phase diagram
of two - dimensional mesoscopic clusters formed
by electrons in a semiconductor quantum dot
or by indirect magnetoexcitons in double quantum dots.
At zero (or sufficiently small) temperature, as quantum fluctuations
of particles increase, two types of quantum disordering phenomena
take place:
first, at small values of quantum de Boer parameter $q < 10^{-2}$
one can observe a transition
from a completely ordered state to that in which different
shells of the cluster, being internally ordered,
are orientationally disordered relative to each other.
At much greater strengths of quantum fluctuations, at $q \sim 0.1$,
the transition to a disordered
(superfluid for the boson system) state takes place.
\end{abstract}

68.65$+$g: Low-dimensional structures: structure, and nonelectronic properties;\\
36.40Ei: Phase transitions in clusters\\

The study of properties of mesoscopic systems is becoming very
important due to the unabated reducing of the characteristic sizes of
electronic devices.
The progress in microlithography and semiconductor technology
have enabled one to perform experiments with ultrasmall structures,
containing as small as several electrons or excitons.
Study of such systems has led to the development of
one - electronics \cite{Ashoori} and have provided essential
progress toward a construction of different mesoscopic systems
(see, e.g., \cite{mesosc}) which can be considered as a base of
nano- and optoelectronics of the future.

In the present paper we consider the phase diagram of
two-dimensional (2D) mesoscopic system of particles interacting through
the Coulomb and dipole potential and confined by a quadratic potential.
These models can be used when describing electrons in a
semiconductor quantum dot \cite{Ashoori,Koulakov} (Coulomb cluster) or
excitons in a system of vertically coupled quantum dots
\cite{mesosc,qdot,Loz} (dipole cluster).
Present experimental technique enables one to prepare the clusters
of a given number of particles in both classical and quantum regimes
\cite{Ashoori,note}.
Such possibility gives one an opportunity to explore a set of
intriguing problems of the physics of mesoscopic systems.

One of the most interesting features of mesoscopic classical
clusters is the existence of the
orientational disordering phenomenon ("orientational melting of
small clusters", see \cite{Mandelstam} - \cite{Belousov}),
when at temperatures many orders less than that of the total melting
(of the disruption of cluster shells) mutual orientational disordering of
different parts of a system takes place,
pairs of shells rotate as a whole relative to each other loosing their
relative orientational order.

When the amplitudes of quantum fluctuations
are comparable with the mean interparticle distance
one can raise the question about the existence of a quantum analog
of the classical orientational melting phenomena.
One can expect that an increase in the strength of quantum fluctuations
should lead to the lowering of orientational melting temperature
$T_{s_1 s_2}$ of a pair of shells $\{s_1, s_2\}$ and,
to all appearances, at some critical strength of quantum fluctuations
"zero - point" orientational melting should take place.

What defines the position of the point $q_{s_1 s_2}$ of
zero-temperature orientational melting?
How will be arranged the regions in which different pairs of a
multishell cluster are orientationally ordered?
What is the role of the statistics in this phenomena?
In the present work we try to answer this questions with the help of
the path integral Monte Carlo calculations.

Mesoscopic clusters, properties of which we are studying, can be
considered as two-dimensional system of $N$ particles of mass $m$
in a parabolic confinement potential of strength $\alpha$.
The general form of the corresponding Hamiltonian is:
\begin{eqnarray}
\hat H = \frac{\hbar^2}{2m} \sum\limits_{i=1}^{N}
\left( \frac{\partial^2}{\partial x_i^2} +
\frac{\partial^2}{\partial y_i^2}\right) +
\alpha \sum\limits_{i=1}^{N} \left( x_i^2 + y_i^2 \right) +
\sum\limits_{i<j} U(r_{ij}),
\quad r_{ij} = |{\bf r}_i - {\bf r}_j|
\label{H1}
\end{eqnarray}
The forms of the interaction potential considered
in this Letter are {\bf 1)} the Coulomb interaction of electrons $e$ in a
semiconductor quantum dot
$U(r_{ij}) = e^2 / r_{ij}$ and {\bf 2)} the dipole interaction
$U(r_{ij}) = d^2 / r_{ij}^3$.
The latter interaction potential corresponds, {\it e.g.}, to the system of
indirect magnetoexcitons in a vertically coupled double quantum dot \cite{Loz}
or to electron system near metal gate that modifies
interelectron interaction potential due to polarization \cite{Yudson}.
The dipole moment $d=he$ is defined by the distance $h$ between the dots.

Hamiltonian (\ref{H1}) can be reduced to a dimensionless form
if all distances and energies are expressed in the units of
$r_0$ and $E_0 = \alpha r_0^2$,
where $r_0 = e^{2/3} / \alpha^{1/3}$ for a system of electrons
(when the Coulomb cluster is considered; $Cul$);
$r_0 = d^{2/5} / \alpha^{1/5}$ for a cluster of magnetoexcitons
(dipole cluster; $D$).
Within the above units (\ref{H1}) has the form:
\begin{eqnarray}
\label{H}
\hat H = q^2 \sum\limits_{i=1}^{N}
\left( \frac{\partial^2}{\partial x_i^2} +
\frac{\partial^2}{\partial y_i^2}\right) +
\sum\limits_{i=1}^{N} \left( x_i^2 + y_i^2 \right) +
\sum\limits_{i<j} U(r_{ij}),\\
\nonumber
U(r_{ij}) = \left\{ 1 / r_{ij}, Cul; \quad
                    1 / r_{ij}^3, D \right\}
\end{eqnarray}
The strength of quantum fluctuations is controlled by
the parameter
$q = \hbar / \left( m^{1/2} \alpha^{1/2} r_0^2 \right)$.
The other dimensionless parameter which defines the state of the system
(\ref{H}) is the dimensionless temperature
$T = k_b T / \left( \alpha r_0^2 \right)$.

In the path integral Monte Carlo approach
(the multilevel block algorithm \cite{Ceperley} have been used)
the properties of a given 2D quantum system
$\{ \hat{\bf r}_i \}, i=\overline{1..N}$
are estimated with the help of a fictitious 3D one
$\{ {\bf r}_i^p \}, p=\overline{0..P-1}$
resulting from the discretisation of functional integrals.
The required accuracy of this substitution, being controlled by the
dimensionless parameter $\tau = q / (P T)$,
have been achieved by adjusting the number of levels of 3D system in such a
way as to best satisfy the condition $\tau = 0.3$.

The state of the cluster at each considered point $\{q;T\}$
of dimensionless control parameters
have been inferred from the analysis of the radial distribution function,
radial mean-squared displacements of particles $u^2_r$.
Orientational disordering of a cluster have been tested with the help of
the mutual orientational order parameter $g_{s_1 s_2}$ \cite{Belousov}.
We have considered also mean - squared radial $l_r$ and angular
$l_{\varphi}$ fluctuations of trajectories in imaginary time,
i.e. the measure of  quantum "librations" of
particles \cite{Mandelstam,librations}.

Let us consider first the behavior of classical clusters ($q=0$).
The ground state configuration (at $T \ll 1$) of a classical dipole
system of 10 particles in a parabolic confinement has two distinct
shells, the inner of which contains 3 particles.
The corresponding configuration can be designated as $D_{10}(3,7)$.
The insert of Fig.~1 shows that an increase in the temperature leads first,
at $T = T_{21} \approx 5.4 \cdot 10^{-6}$, to the
disappearance of the mutual orientational order of shells.
Total disordering in the system, when radial shell order disappears,
takes place at much more high temperature $T_f \approx 0.01$.

It is obvious that in small clusters consisting of several shells
a set of orientational melting phenomena will take place,
each of them will appropriate to relative orientational disordering of
different pairs of shells.
This peculiarity of small clusters is well observed in Fig.~1 which
represents temperature dependencies of radial fluctuations
$u^2_r$ and of mutual orientational parameter $g_{s_1 s_2}$
of three - shell Coulomb cluster $Cul_{25}$
the ground state configuration of which can be written as $Cul_{25}(3,9,13)$.
Note, that the data plotted in the figure reinforce the statement that
the temperatures $T_{s_1 s_2}$ of the orientational "melting"
of the pair of shells $\{s_1, s_2\}$ are functions of the particle
distribution throughout shells $(N_{s_1},N_{s_2})$ and are maximal
when the shells involved are parts of ideal 2D hexagonal
crystal (i.e. when  $(N_{s_1},N_{s_2}) = (3,9), (4,10), (6,12), ...$).
\cite{Bedanov,Belousov}

As it follows from the Fig.~1, the temperature interval
$T_{s_1 s_2} < T < T_f$
can be considered as a region in which shells $\{s_1,s_2\}$
of a cluster rotates relative to each other retaining their internal order
and loosing the relative one.
In $2 N$ - dimensional configurational space the system moves then
along a narrow ravine on a potential energy surface,
different points of this ravine correspond to different values of
the mutual order parameter $g_{s_1 s_2}$.
Fig.~1 shows that this movement is accompanied by a "breathing"
of cluster shells to lead to sharp increase in the value of radial
mean - squared fluctuations $u^2_r$ at the points $T_{s_1 s_2}$
of orientational disordering.

Let us consider how changes the state of clusters
as the strength of quantum fluctuations is increased at constant temperature.
Shown in Fig.~2a is the behavior of mutual orientational order parameter
$g_{21}$ of dipole cluster $D_{10}$ when the system is moved along line
$T=3 \cdot 10^{-6}$.
A sharp variation in the value of $g_{21}$
at the point $q_{21} \approx 8 \cdot 10^{-4}$ testifies about the
quantum fluctuations induced transition from the orientationally ordered
(OO, at $q < q_{21}$) to the orientationally disordered but radially ordered
state (RO, at $q > q_{21}$).

Analogous results take place for Coulomb cluster $Cul_{25}$
and are plotted in the insert of Fig.~3a.
The results shown in the figure correspond to the temperature
$T = 3 \cdot 10^{-4}$ at which the orientational order of the
first and the second shells of the cluster breaks down at $q \approx 0.01$.
Quantum orientational melting of the second and the third shells
takes place at sufficiently lower values of quantum parameter:
$q_{32} \approx 8 \cdot 10^{-3}$.

The dependencies of angular $l_{\varphi}$ and radial $l_r$
quantum librations of particles as functions of quantum parameter $q$ is shown
in Fig.~2b.
In the insert of this figure a typical picture of imaginary - time
trajectories (of their projections onto OXY plane)
of the cluster is presented.
The picture shows that quantum fluctuations are strongly anisotropic.
An increase in the value of quantum parameter leads to a significant
increase in the angular fluctuations at the point $q_{21}$ of
disappearance of an orientational order, while radial fluctuations
$l_r$ do not have any peculiarities in the region considered.

Above results show that at the point of
quantum orientational melting $q_{s_1 s_2}$
the characteristic scale of quantum motion becomes comparable with the
mean angle between nearest particles of the pair of shells $\{s_1, s_2\}$.
A special feature of this type of quantum "melting" in mesoscopic systems
is that at low temperatures main directions of particles fluctuations
in corresponding classical system are defined by a narrow and high
ravine formed by a potential energy surface.
The presence of such ravine leads to a strong anisotropy of quantum
fluctuations and sets off relative "quantum rotations" of a given
pair of shells.
It is also worth to note that orientational melting in
both Coulomb and dipole clusters take place in that region of dimensionless
parameters $\{q;T\}$, in which the role of quantum statistics is
unimportant (see the insert of Fig.~2b).

Let us consider concisely how behave mesoscopic systems as the strength
of quantum fluctuations is increased further, at $q \ge 0.05$.
The simplest estimations show (see below) that in this region
of phase diagram, as $q > \sqrt{T}$, quantum exchanges of particles
became to play an important role.
In order to simplify our calculations, we have supposed the particles
of the cluster to be bosons.
Strictly speaking, in our case this can be applicable only for the
dipole system of indirect magnetoexcitons in double quantum dots \cite{Loz}.
Electrons in a semiconductor dot obeys Fermi statistics and the model
used collapses in the region of very strong quantum fluctuation.
Meanwhile, the study of Coulomb boson system is of some methodical
interest offering a clearer view of common features and regularities
in behavior of boson clusters.

In Fig.~3a the results of calculations of mean-squared radial fluctuations
$u^2_r$ in dipole and Coulomb boson clusters at $T = 5 \cdot 10^{-3}$
are plotted {\it vs.} quantum parameter $q$.
As it follows from the above analysis (see Fig.~1-2),
in the considered region of control parameters all clusters are in
orientationally disordered state.
Therefore, a sharp increase in the value of $u^2_r$ testifies about the
total "melting" in these systems.
A comparison between the behavior of radial fluctuations (Fig.~3a)
and that of the superfluid fraction \cite{note_ns} (Fig.~3b)
shows that this disordering corresponds to the transition from the
radial ordered (RO) to the superfluid (SF) state.
By way of illustration, in the insert of Fig.~3a we have plotted some typical
pictures of imaginary - time trajectories of the system.
Cyclic permutations of particles are well noticeable above the transition.
As temperature $T_f$ of the total disordering in appropriate
classical systems is $T_f \approx 0.01$ and is approximately two times
higher of the temperature to which the data in Fig.~3 belong,
values $g_f$ of the total disordering can be taken as rather
good estimations of locations of quantum phase transitions (at $T=0$).

By combining results considered above one can sketch the
"phase diagram" of small quantum clusters on plane $\{q;T\}$, see Fig.~4.
Of course, considering systems of such a small number of particles
it is impossible to speak about the existence of distinct
boundaries between states cited above, treating them as
lines $T_c(q)$ of phase transitions.
But, as one can see from Fig.~1-3, an analysis of the system of even
such a small number of particles enables one to reveal the presence of such
regions as well as its arrangement in plane $\{q;T\}$.

The mutual arrangement of different regions of the
phase diagram is controlled by the following relations:
{\bf 1)} At small strength of quantum fluctuations the ratio of the
temperature to the energy barriers
for relative rotations of some pair of shells $\{s_1 s_2\}$
defines the region
of orientational disordering $T > T_{s_1 s_2}$ of this pair of shells.
{\bf 2)} Subsequent increase in the temperature (keeping $q = const.$)
leads to the transition to the "classical liquid" (CL)
state at $T > T_f$
with the temperature $T_f$ controlled by the ratio of the temperature
of the system to the characteristic energy of interparticle interaction
(i.e. by dimensionless temperature $T$ in our units).
{\bf 3)} At $T > T_f$, as the strength of quantum fluctuations is increased,
and the thermal de Broigle wavelength becomes comparable with the
mean interparticle distance in the liquid,
the liquid / superfluid transition takes place along line $T(q) = q^2$.
{\bf 4)} Zero - temperature point $q_f$ of a quantum melting,
the transition of a cluster from a radial ordered to
superfluid (SF) state, is controlled by the ratio between the energy
of quantum fluctuations and the characteristic energy
of interparticle interaction.
{\bf 5)} Comparing the energy of quantum fluctuations of particles with the
energy barrier for relative rotations of pair $\{s_1,s_2\}$
of cluster shells enables one to estimate the position of the quantum
orientational melting point $g_{s_1 s_2}$.

In conlusion, we note that the phenomena of quantum orientational melting
can also take place near rare impurities of an infinite system, as
in the vicinity of an impurity a shell structure can arise,
resembling one of an isolated cluster.

The work have been partially supported by Russian Foundation of
Basic Researches, INTAS and by Sweden Academy of Science.


Figure~1 \\
Classical three - shell Coulomb cluster $Cul_{25}$, $q=0$.
The temperatures $T_{s_1 s_2}$ of orientational melting of shells
$\{s_1;s_2\}$ are marked by arrows.
Shown also the temperature $T_f$ of the total melting in the system,
the transition from a radial ordered (RO) to a "classical liquid" (CL)
state. \\
Insert: Classical two - shell dipole cluster $D_{10}$, $q=0$.

\vspace{0.3cm}

Figure~2 \\
Quantum orientational melting of dipole cluster $D_{10}$
at $T = 3 \cdot 10^{-6}$ and of Coulomb system $Cul_{25}$ (in the insert
of the Fig.~3a) at $T = 5 \cdot 10^{-3}$. \\
a) Mutual orientational order parameter $g_{21}$ {\it vs.} quantum
parameter $q$.\\
b) A measure of quantum "librations" of particles:
radial and angular variances of particles trajectories in imaginary time
$l_{r}(q)$ and $l_{\varphi}(q)$.
The region of abrupt changes in $l_{\varphi}$ coincides with that of
the disappearance of orientational order.\\
Insert: a typical picture of projections of imaginary - time trajectories
onto OXY plane (the picture is appropriate to cluster $D_{10}$
at the point $\{q;T\} = \{ 8\cdot 10^{-4}; 3\cdot 10^{-6} \}$).

\vspace{0.3cm}

Figure.~3 \\
$T = 5 \cdot 10^{-3}$. \\
a) Radial fluctuations $u^2_r$ {\it vs.} quantum parameter $q$.
Inserts present spline - interpolations of OXY-projections of
imaginary - time trajectories just
before ($q = 0.13$) and after ($q = 0.2$) the OO/RO transition. \\
b) Superfluid fraction $\nu_s$ {\it vs.} $q$.

\vspace{0.3cm}

Figure.~4 \\
The "phase diagram" of mesoscopic clusters (all notations are
given in the text).
Dashed (solid) lines correspond to dipole cluster $D_{10}$
(Coulomb system $Cul_{25}$).
The region of transitions from orientationally ordered to
orientationally disordered states is shown qualitatively on an enlarged scale.


\begin{thebibliography}{99}
\bibitem{note} the strength of quantum fluctuations in the system of excitons
in double quantum dot can be controlled by the applied magnetic field $H$,
as the (magnetic) mass of the exciton $m$ can be shown to
depend on $H$ as $m(H) \sim \sqrt{H}$ in the limit of low dense system
\protect\cite{Loz}.
\bibitem{note_ns} the superfluid fraction
$\nu_s$ have been measured through the response of the system to its
infinitesimally slow rotation \cite{Ceperley}.
\end{thebibliography}
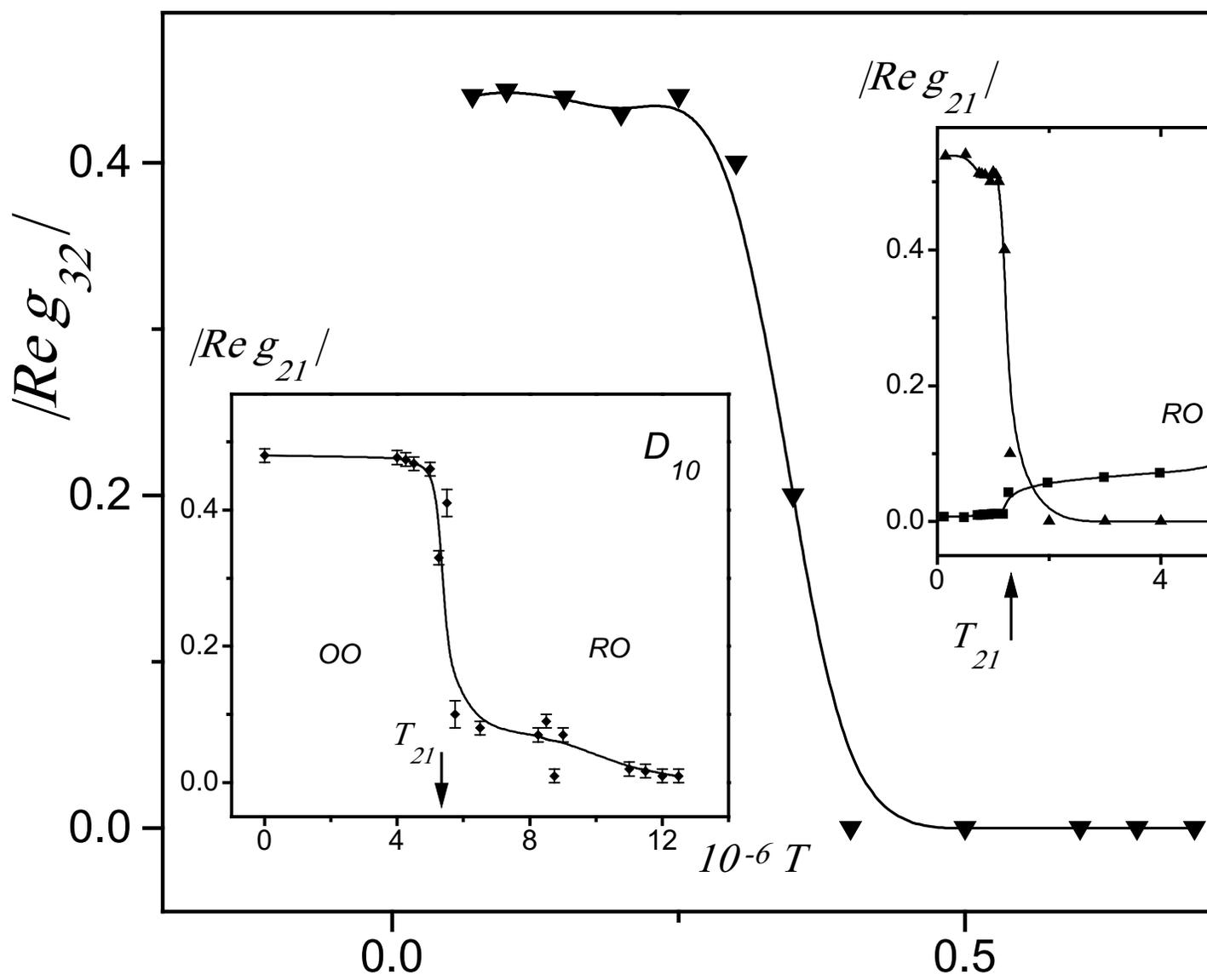
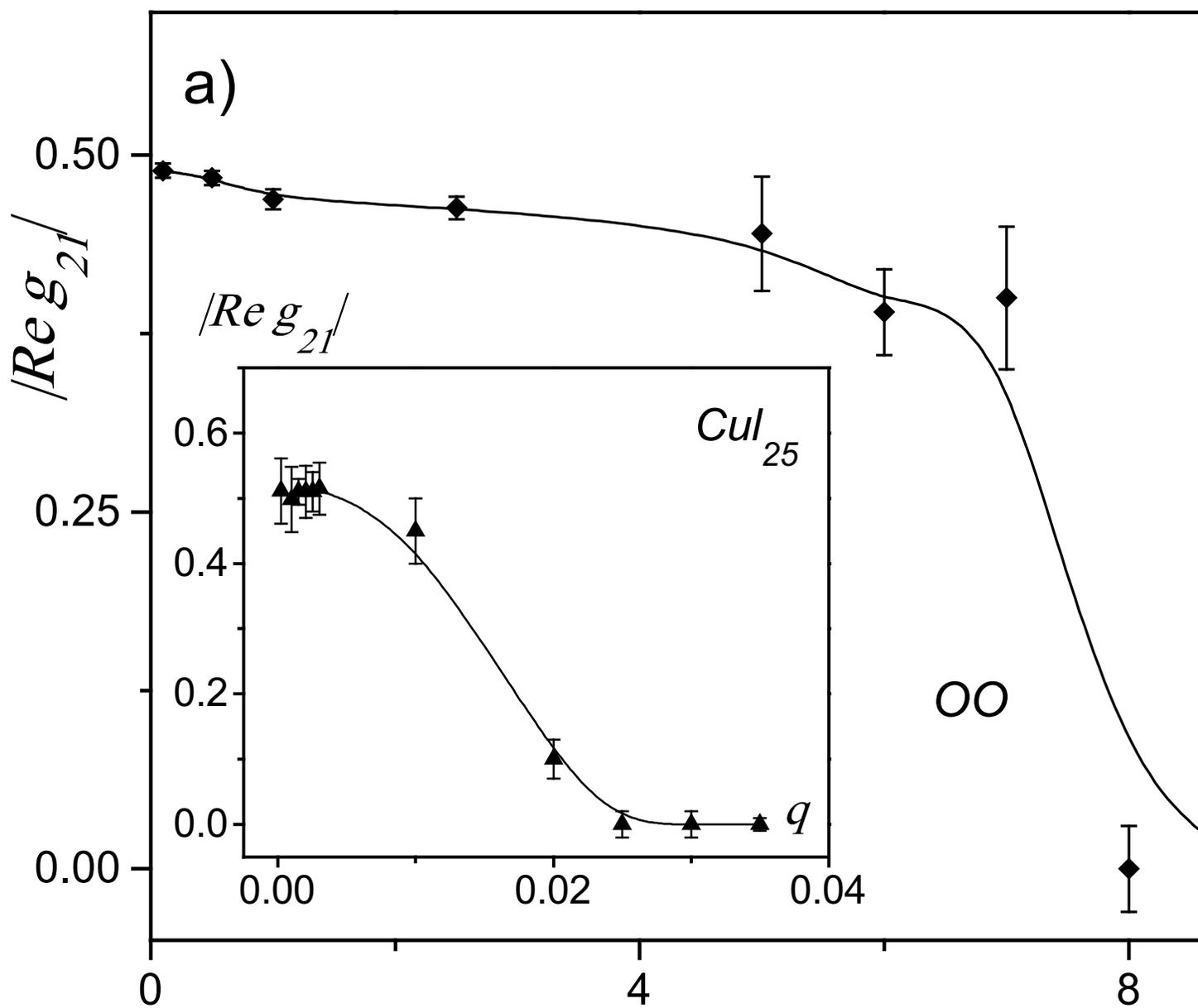
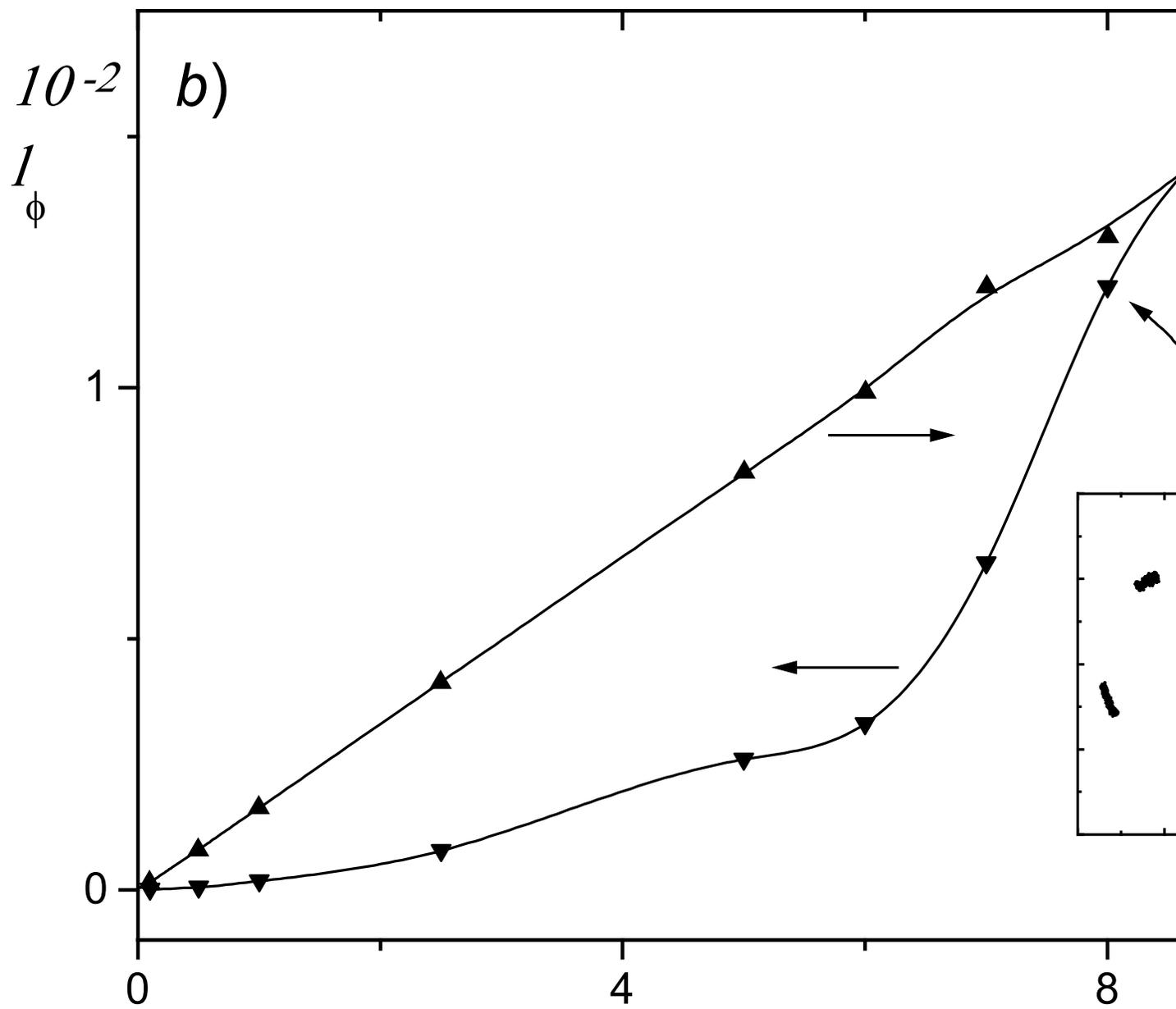
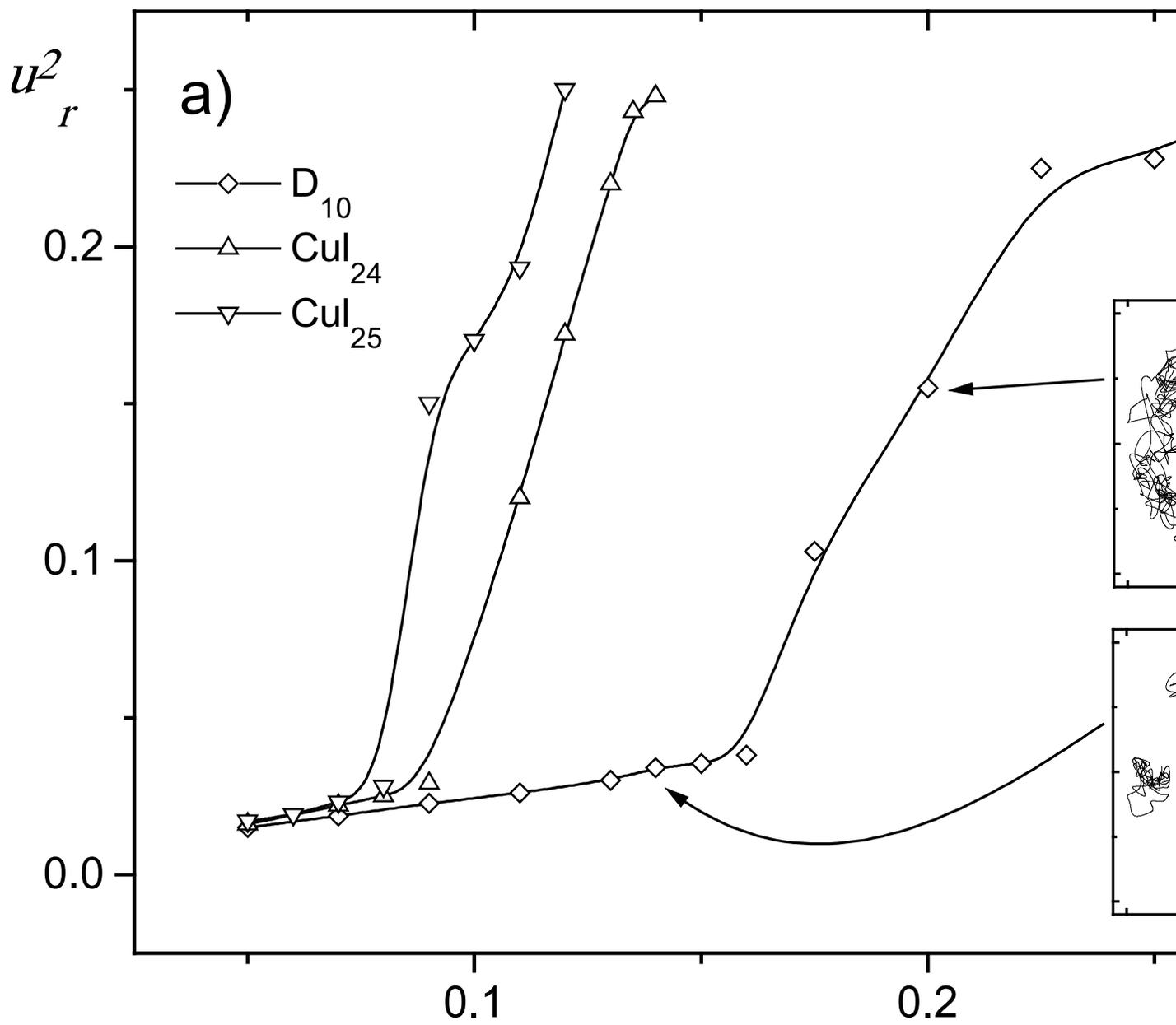
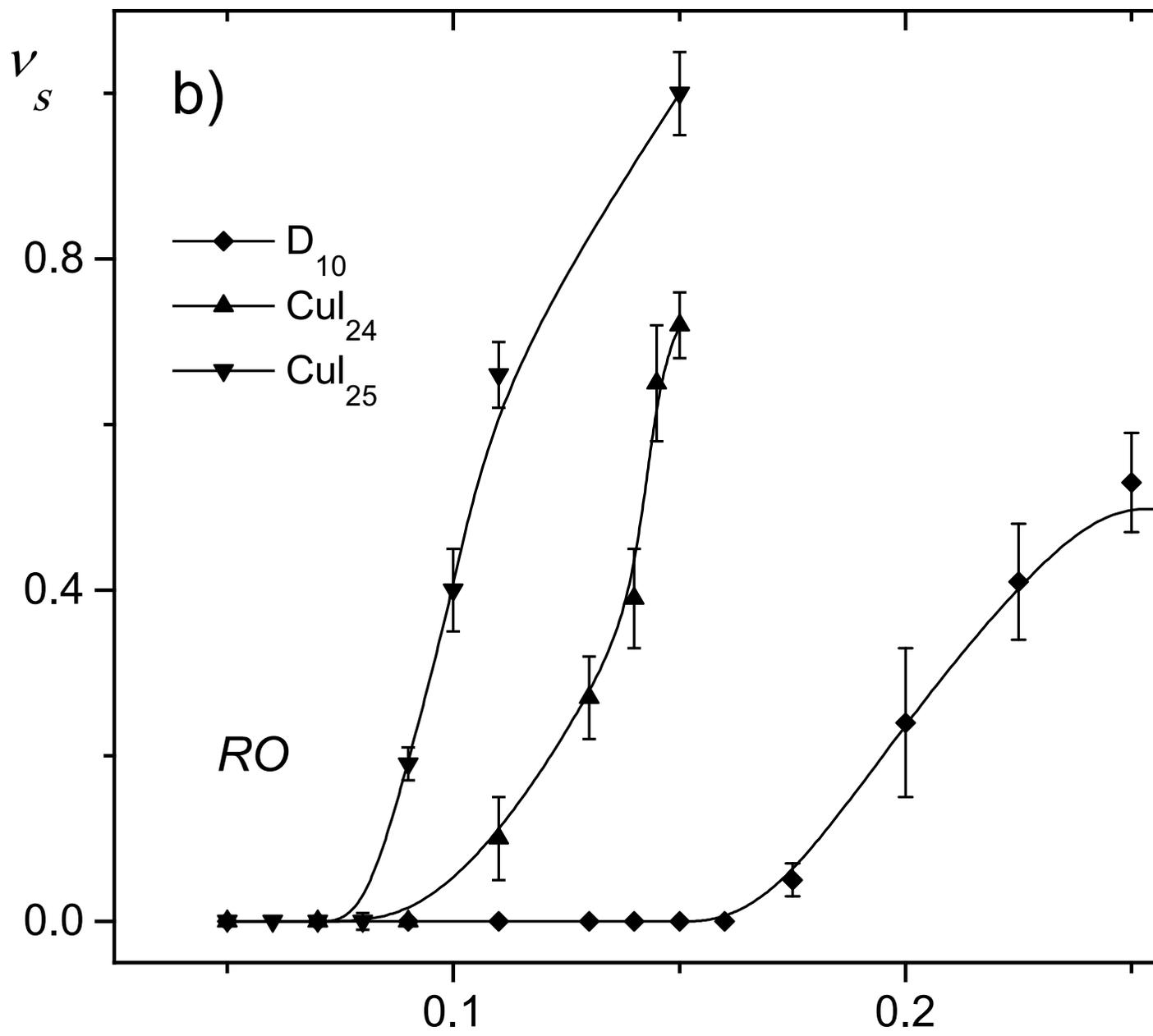
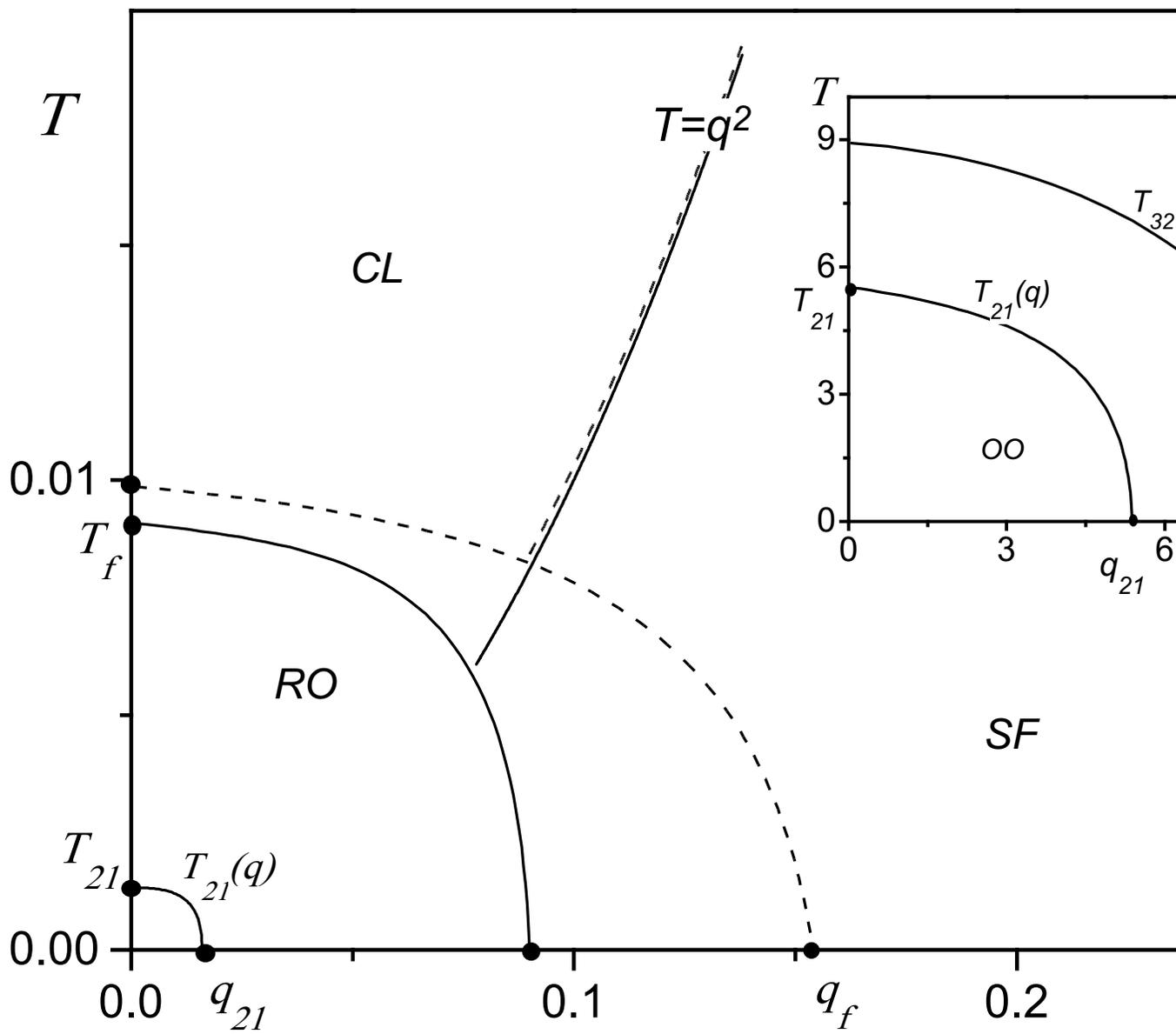
\end{document}